\documentclass[twocolumn]{pasj00}

\SetRunningHead{H. Sudou et al.}{VLBI Observations of IRC+60169}

\title{VLBI Observations of Water Masers in the Circumstellar Envelope
of IRC+60169} 
       
\author{Hiroshi \textsc{Sudou},$^1$
       Toshihiro   \textsc{Omodaka},$^2$
       Hiroshi   \textsc{Imai}, $^{3,4,}$
       \thanks{Present address: Joint Institute for VLBI in Europe,
       Postbus 2, 7990 AA Dwingeloo, the Netherlands}
       Tetsuo   \textsc{Sasao},$^4$\\
       Hiroshi   \textsc{Takaba},$^{5,6}$
       Masanori   \textsc{Nishio},$^2$
       Wataru   \textsc{Hasegawa},$^2$
       and Junichi   \textsc{Nakajima}$^6$\\
       }
       \altaffiltext{1}{
       Astronomical Institute, Graduate School of Science,
       Tohoku University, Aoba, Sendai 980-8578}
       \email{sudou@astr.tohoku.ac.jp}
       \altaffiltext{2}{
       Faculty of Science, Kagoshima University, 1-21-30 Korimoto,
       Kagoshima 890-0065}
       \altaffiltext{3}{
       Mizusawa Astrogeodynamics Observatory, National Astronomical
       Observatory, Mizusawa, Iwate 023-0861  }
       \altaffiltext{4}{
       VERA Project office, National Astronomical Observatory,
	2-21-1 Osawa, Mitaka, Tokyo 181-8588}
       \altaffiltext{5}{
       Faculty of Engineering, Gifu University,
       1-1 Yanagido, Gifu 501-1193}
       \altaffiltext{6}{
       Kashima Space Research Center, Communications Research
       Laboratory, Kashima, Ibaraki 314-0012}
       
\KeyWords{
masers ---
stars: circumstellar matter --- 
stars: late-type ---
stars: individual (IRC+60169) ---
techniques: interferometric}
\Received{2002 May 7}
\Accepted{2002 August 3}
\Published{}

\begin{document}
\maketitle

\begin{abstract}
Water masers around an AGB star, IRC+60169, were 
observed at four epochs using the Japanese VLBI networks. The
distribution of the maser 
features is limited in a thick-shell region, which has inner 
and outer expansion velocities of 7 km s$^{-1}$ and 14 km s$^{-1}$ at 
radii of 25 mas and 120 mas, 
respectively. The distribution of the red-shifted 
features exhibits a ring-like structure, the diameter 
of which is 30 mas, and corresponds to the inner radius 
of the maser shell. This implies that dense gas around 
the star obscures red-shifted emission. Although a position--radial
velocity diagram for the maser features is consistent with a spherical
shell  
model, the relative proper motions do not indicate an expansion  motion
of the shell. A remarkable property  
has been found that is a possible periodic change of the
alignment pattern of water maser spots.  
\end{abstract}

\thispagestyle{headings}

\section{Introduction}
A low-mass star passes the Asymptotic Giant Branch (AGB) phase in the HR
diagram, in which it undergoes heavy mass-loss processes. After the
AGB phase, the star forms a planetary nebula (PN) by ionizing its
circumstellar envelope (e.g., Kwok 1993).  It has been stated that many 
young PNe exhibit a bipolar morphology or large deviation from spherical
symmetry 
(e.g., Aaquist, Kwok 1991; Sahai et al. 1998). There still exists a
missing link in the mass-loss history between AGB stars and
central stars of PNe.  

Most AGB stars show maser emission from water molecules (Reid,
Moran 1981; Elitzer 1992). Masers have been used for probes of the
mass-loss process in the circumstellar envelope of the AGB star.
VLBI (Very Long Baseline Interferometry) observations
have revealed that relatively young AGB stars (i.e., Mira variables and
semi-regular variables) already exhibit bipolarity in
their spherically expanding flows (e.g., Marvel 1997; Ishitsuka
et al. 2001). On the other hand, for more
evolved AGB 
stars (i.e., IRC/AFGL objects and OH/IR stars) and proto-PNe,  little is
known about the detailed mass-loss process by using the 
maser kinematics (cf., Marvel, Boboltz 1999; Imai et al. 2002). 

IRC+60169 has been classified as an IRC/AFGL object using the IRAS
two-color diagram (van der Veen, Habing 1988; Takaba et al. 1994). 
It shows a clear double-peaked spectrum of water masers, which is a
typical signature of IRC/AFGL objects and OH/IR stars. The shape of
the spectrum 
exhibited a violent time variation. In 1977 the
red-shifted component was stronger than the blue-shifted one, and in 1982 an
intensity change between these components occurred (Engels et
al. 1988). This
fact is thought to be related to the temporary variation in the
mass-loss flow. 

We present the water-maser observations of IRC+60169 using Japanese
VLBI networks in order to reveal the origin of the complicated
time-variation in the intensity and distribution of the water masers and to
investigate the kinematics of the circumstellar envelope by measuring
the proper motions of the masers.

\section{Observations}
The water masers of IRC+60169 were observed using KNIFE
(Kashima--Nobeyama InterFErometer; e.g.,
Miyoshi et al. 1993) 
on 1992 June 3 and J-Net (the Japanese domestic VLBI Network; e.g.,
Omodaka et al. 1994)  on 1997
March 30, 1997  
May 17, and 1998 April 4.  The KNIFE system consists of two telescopes
at Kashima and Nobeyama; its minimum fringe spacing is 14 mas. The  
J-Net system consists of four telescopes at Kashima, Nobeyama, 
Mizusawa, and  Kagoshima; its minimum fringe
spacing is 2 mas.  The telescopes used for each observation
are summarized in table 1. 
Since the quality of the data of the 2nd
epoch taken at the Kagoshima telescope appeared to be
poorer than those taken at the other telescopes, we flagged out
the visibilities of baselines including this telescope. 
The received signals were recorded using the K-4 
back-end system, which has 16 video channels with a
band width of 2 MHz each (Kiuchi et al. 1997).
Correlation 
processes have been carried out using the FX correlator at Mitaka
(Chikada et al. 1991).
Each video band was divided into 256 frequency channels, which
corresponds to a velocity 
resolution of 0.11 km s$^{-1}$.

We used the NRAO AIPS package for fringe
fitting and image synthesis. The extragalactic continuum source DA 193 was
used as a calibrator of instrumental clock delays. Successive velocity
channels containing 
the strong maser spots were used to calibrate the fringe
rates. The same channels were also used as fringe-phase references for a
self-calibration procedure. 
The absolute intensities at 
the 2nd, 3rd, and 4th epochs were not measured correctly because of  a
problem of a measurement of either system temperatures or conversion factors.
The beam size of the KNIFE observation was 
about 14 mas with an almost circular shape, and those of the J-Net
observations were 5.9$\times$3.6, 8.9$\times$4.7, and 6.3$\times3.7$
[mas$\times$mas], at the 2nd, 3rd, and 4th epochs, respectively.

Maser emission is spatially resolved  
into some ``maser features'', each of which consists of ``maser spots'' with
almost similar 
positions and velocities (within $\sim$5 mas and $\sim$1 km s$^{-1}$,
respectively).  The definitions of the words ``feature'' and ``spot'' 
are the same  as those of Gwinn (1994) and  Imai et al. (2000). The
positions of the maser spots were 
estimated by using the AIPS task 
``JMFIT'', which is used for 2-dimensional Gaussian fits.
We adopted the flux-weighted mean of the positions and the velocities of
the maser spots as the position and velocity of the maser
feature. 

\section{Results}
Figure 1 is a spectral profile of water masers  in IRC+60169
at the  1st epoch.  They were detected over a range of 
between $-10$ km s$^{-1}$ and $-35$ km s$^{-1}$.  The intensity of the
component near $-33$ km s$^{-1}$ is the strongest 
in the spectrum. These characteristics are consistent with recent
observations of this object (e.g., Colomer et al. 2000). Figure 2 shows
the spatial distributions of the water-maser features at all four
epochs. Tables 2--5 give parameters of the
detected maser features at the individual epochs. We 
grouped them into two velocity components: 
blue-shifted ($\leq V_{\rm sys}$) and red-shifted ($\geq V_{\rm sys}$),
where $V_{\rm sys}$ is the stellar systemic  
velocity, which is $-24$ km s$^{-1}$, given by the central velocity of the
SiO-maser spectra (Engels et al. 1988). Hereafter, they are abbreviated  
as the B 
and R components, respectively. As shown in figures 2b, c, and d, the B
component seems to be located in a limited region and 
the R component appears to be
distributed in a ring-like 
region around the B component. 

Figure 3 shows a plot of the relative velocities of the maser features with
respect to the stellar systemic 
velocity against the projected 
radial distance from a reference point. The reference point is the
position of the 
brightest maser feature in each epoch.
Assuming the standard thin-shell approximation, we obtained 
an inner shell radius $R_{\rm in}$ of $\sim$ 25 mas (17 AU, assuming the
distance of 690 pc; Loup et al. 1993) and an 
outer
shell radius $R_{\rm out}$ of $\sim$ 120 mas (83 AU). The expansion 
velocities at these radii are $\sim$ 7 km s$^{-1}$ and $\sim$ 14
km s$^{-1}$, respectively.

Comparing the positions and the velocities among the 2nd, 3rd, and 4th
epochs, we identified eleven common  
maser features, which are labeled {\sf a, b, c, d, e, f, g, h, i, j, {\rm
and} k} in tables 3--5. We did not use the data of the 1st epoch 
because of a too-long  time span ($\sim 8$ yr) to the next
epoch. However, the maser distribution at the 1st epoch is not very different
from those at the other epochs. 
The velocity of each maser feature was almost equal  
within 1 km s$^{-1}$ among the epochs.
Although each of two maser
features, {\sf b} and {\sf d}, had two identifiable features between
the 2nd and 3rd epochs, we  
traced the maser feature 
exhibiting more similar alignment patterns of maser spots.
Features
{\sf a, b, {\rm and} j} were detected 
at the 2nd, 3rd, and 4th epochs. Feature {\sf k} was detected at the
2nd and 4th epochs. The other features were detected at the 2nd and 3rd
epochs. 

The proper motions of the maser features are shown in figure 5 and their
parameters are summarized in table 6. The proper
motions were measured relative to feature {\sf a}, which was detected
at the 2nd, 3rd, and 
4th epochs. For 
features {\sf b} and {\sf j}, a weighted root-mean-square fit was carried
out to obtain the proper motion during the three epochs. Figure 
4 shows the fitting results for features {\sf b} and {\sf j}.
The proper motions do not show a systematic  
motion that is expected from a spherically expanding shell model. 

\section{Discussion}

\subsection{Blocking Maser Emission by a Dense Circumstellar Envelope}
In the case of Mira
variables, red-shifted and blue-shifted components 
are always located in a small 
region, while intermediate  components are always in a
ring-like region. These components indicate radially and tangentially
beamed masers, respectively (e.g., Lane et
al. 1987).  In order to obtain the averaged distribution among
several years, 
we made a superposition of the distribution of the maser
components of IRC+60169 with respect to the
brightest component. 
Since the VLA observation of IRC+60169 reported by
Colomer et al. (2000) showed the same morphology as ours, it was also
added in this superposition. Figure 6a shows that the B 
components appeared to be in a limited region, such as blue-shifted
components around Mira variables. On the other hand, 
in figure 6b, the R component appears
to be in the 
ring-like region around the B component. This morphology can be
explained by a blocking 
effect proposed by Takaba et al. (1994): radially-beamed masers
behind the blocking gas around the star are completely 
obscured, and all of the red-shifted masers widely distributed are 
the tangentially-beamed masers. Note that the R component, which is
likely to be weakly beamed, is weaker than the B component. Engels et
al. (1997) suggested an
anti-correlation of intensities between the radial and tangential
masers. This could be the reason why the intensity change between the
B and the R components of IRC+60169
occurred in 1982. 

The size of the obscuration was roughly estimated to be
$\sim$30 mas (21 AU), indicated from the 
diameter of the apparent absence of the R component. This value is 
in good agreement with the inner radius of the maser shell in figure 3
($\sim 17$ AU). The blocking material is the dense gas with 
a density of hydrogen molecules higher than 10$^{10}$ cm$^{-3}$ 
and quenches water maser emission
(Cooke, Elitzur 1985). This dense-gas region corresponds to the region
where SiO-maser emission occurs, or the radio photosphere where H$^-$ and
H$^-_2$ free--free absorption occurs (Reid, Menten 1997).

A blocking region has not been found in Mira variables. The radius
of such a dense-gas region was
found 
to be a few AU in Mira variables (e.g., Reid, Menten 1997). This value is too
small to block  
water maser emission. However, the radius of the envelope is believed to 
increase with increasing luminosity or the
mass-loss rate. Thus, the blocking region becomes larger when
AGB stars evolve more. Although the luminosity has not been measured,
the mass-loss rate of  
IRC+60169 was estimated to be an order of magnitude larger than that of
typical Mira 
variables (Loup et al. 1993). This fact supports the idea that the
blocking effect is significant in the envelope of IRC+60169

\subsection{Environment of the Maser Shell in IRC+60169}
The outer shell radius (83 AU) estimated from figure 3 is larger than
that obtained by Colomer et 
al. (2000), because they were not able to detect the red-shifted
component existing in the outer region of the shell. 
Although it is still smaller than a typical value for OH/IR 
stars or supergiants ($\sim$100 AU, e.g., Rosen et al. 1978; Yates,
Cohen 1994), it is 
larger than that for Mira variables ($\sim$10 AU, e.g., Spencer et
al. 1979; Bowers, Johnston 1994).  
One more important parameter of the maser shell
is the logarithmic velocity gradient that describes the acceleration,
$\epsilon=d(\ln V)/d(\ln R)$, where $V$ is the expansion velocity and
$R$ is the radius of the shell. For the smaller value of  $\epsilon$, if 
the coherence path
into the radial direction becomes longer,  radial beaming occurs. In
contrast, for a large value of  $\epsilon$, if the
coherence path
into the tangential direction becomes longer,  tangential beaming occurs. 
Using the obtained parameters in the spherically expanding shell model
for IRC+60169, we find  
$\epsilon \sim 0.4$.
It is smaller than those obtained in Mira variables and semi-regular variables
(Ishitsuka et al. 2001). This fact suggests that the radial masers are more
dominant in IRC+60169 than Mira variables. Therefore, the
radially-beamed, red-shifted  masers are 
easy to suffer the blocking effect. 

\subsection{Alignment Patterns of Maser Spots}
It is stated that maser features exhibit a linear or curved maser-spots
alignment in some supergiants
(e.g., Richards et al. 1996, 1998) and some star-forming
regions (e.g., Gwinn 1994; Torrelles et al. 2001), which
are thought to be caused by  shocks. 

We focus our attention on the time variation of 
the maser-spots complex, including feature {\sf j} and nearby
features.  Figure
7a shows the spatial distribution of maser spots consisting of
individual features.   
The alignment pattern of the 
maser spots has changed between the 2nd and 3rd epochs, while
at the 4th epoch, 
it appeared to be the same pattern as that at the 2nd epoch. The same 
tendency can be seen in the velocity distribution (figures
7b and c).

The alignment change is
thought to be periodical and  dominated by the
pulsation of the star, 
because the time span between the 2nd and 4th epoch (360
d) is close to the IR variability period of IRC+60169 (440$\pm$30 d, 
Lockwood 1985). This is explained by an analogy of the
mode-switching effect proposed by Engels et al. (1997).
The radius at which water masers occur could change according to the
stellar brightness.
If this hypothesis is correct, the proper motions of 
the maser features are a mixture of the bulk and pattern
motions. This fact gives us difficulties in    
identifying the maser features and interpreting the proper motions. 

\subsection{Distance Measurement}
Assuming an isotropic and random motion, 
we are able to estimate the distance to a star by applying the statistical
parallax method (e.g., Genzel et al. 1981). 
Although the proper motion of the maser features in IRC+60169 might
include the pattern motion of the maser  
cloud, we ignore this effect in this subsection. 
We have obtained the standard deviation of the Doppler velocities of 7.3
km s$^{-1}$ and    
those of the proper motions in the R.A. direction and the
Dec. directions of 5.7 and 4.4 mas yr$^{-1}$, respectively. 
These values give distance values of 280 $\pm$ 50 pc and 360
$\pm$ 60 
pc, respectively, taking into account the observational
errors.
On the other hand,  
the distance of 690 pc was suggested by assuming that the bolometric luminosity
of IRC+60169 is $10^4 L_{\odot}$ (Loup et al. 1993).  The 
underestimation of the distance could be caused by the blocking effect,
because the radial velocity dispersion 
is underestimated. Assuming that the velocity distribution of the R
component is the same as that of the B component, we obtain a velocity
dispersion of 7.5  km s$^{-1}$, and the distance is estimated
to be 290 and 370 pc, respectively. Further, adopting that the stellar
systemic velocity is $-23$ km s$^{-1}$, suggested from CO observations
(e.g., Lindqvist et al. 1988),  we obtain a velocity
dispersion of 8.4  km s$^{-1}$ and the distance is estimated
to be 320 and 410 pc, respectively. Since these values are still smaller
than the distance obtained by Loup et al. (1993), 
an intrinsic asymmetry of the velocity field of the
envelope could be the reason for the underestimation of the 
distance.

\section{Conclusions}
In the present work,  multi-epoch VLBI observations of the water masers
in IRC+60169 were carried out. The main results from
our data are described as follows:
1) The blocking effect is suggested by the ring-shaped distribution of the
red-shifted maser features. The blocking size was estimated to be
$\sim 20$ 
AU, which almost equals to the inner edge of the maser shell. This
phenomena is believed to be seen only in evolved AGB stars.
2) The maser-shell size ($\sim$80 AU) suggests that IRC+60169 is more evolved
than Mira variables. 
Although random motions are likely to dominate the kinematics of the
envelope, radial acceleration also occurs in the maser shell. 
3) We found that the 
alignment pattern of the maser spots seems to change periodically with
the stellar pulsation. This implies that the spots pattern is 
stationary during a year, and that the environment in which masers occur moves 
to the radial direction.
4) Using the statistical parallax method,
we obtained a weighted-mean distance to IRC+60169 of 310$\pm$40
pc. This value is 
probably underestimated because of the blocking effect and/or the
intrinsic asymmetrical velocity field. 

Since we have measured only eleven proper motions, the kinematics of the
circumstellar envelope in IRC+60169 is still unclear. Further systematic
monitoring observations every few 
months will reveal the kinematics of the envelope, the distance, and the
environment of the maser emitting region. In addition,  SiO maser
observations of IRC+60169 will 
be useful in revealing properties of the blocking gas, because it is
considered to exist in it.

\bigskip
We would like to express our gratitude to each member of the Japanese
VLBI networks for their support during observations, correlations, and
data reduction. We also wish to thank an anonymous referee for useful
comments. 
HI and HS were supported by the
Grant-in-Aid for JSPS Fellows by Ministry of
Education, Culture, Sports, Science  and Technology.



\begin{table}[h]
\begin{center}
\caption{Observing dates and participating telescopes.}
\begin{tabular}{ccl}
\hline\hline
 Epoch &
 Date &
 Telescope\footnotemark[$*$] \\
 \hline
 1st &1992 June 3, 1992 & O, N\\
 2nd &1997 March 30& O, N, M , K\\
 3rd &1997 May 17&O, N, M\\
 4th &1998 April 4& O, M\\
\hline
 
\end{tabular}
\end{center}
\par\noindent
 \footnotemark[$*$] The keys to the telescopes are as follows.\\
 O: 34-m telescope at Kashima.\\
 N: 45-m telescope at Nobeyama.\\
 M: 10-m telescope at Mizusawa.\\ 
 K: 6-m telescope at Kagoshima.
\end{table}


\begin{table}[h]
\small
\begin{center}
\caption{Parameters of the detected maser features at
 the 1st epoch (1992 June). }
 
\begin{tabular*}{\columnwidth}{@{\hspace{\tabcolsep}
\extracolsep{\fill}}ccccc}
\hline\hline 
 {$V_{\rm LSR}$ }&
 $\Delta x$\footnotemark[$*$] & Error &
 $\Delta y$\footnotemark[$\dag$] & Error \\
 
 [km s$^{-1}$] &
 [mas] & [mas] &
 [mas] & [mas] \\
\hline
  $-$14.8 &    $-$34.1 & 0.5 &    $-$39.2 & 0.5 \\
  $-$16.9 &    $-$35.3 & 1.1 &    $-$49.6 & 0.2 \\
  $-$16.4 &      +6.0 & 0.9 &    $-$29.8 & 0.8 \\
  $-$16.4 &      +0.6 & 0.2 &     +19.8 & 0.3 \\
  $-$17.5 &    $-$68.4 & 0.8 &    $-$11.8 & 2.9 \\
  $-$16.8 &    $-$99.0 & 0.6 &     $-$7.5 & 0.7 \\
  $-$18.6 &    $-$32.5 & 0.6 &    $-$51.0 & 0.6 \\
  $-$18.0 &    $-$25.7 & 0.2 &    $-$34.1 & 0.2 \\
  $-$18.1 &    $-$17.3 & 0.5 &     +14.7 & 0.9 \\
  $-$18.0 &    $-$89.2 & 0.5 &     +29.9 & 1.3 \\
  $-$18.2 &    $-$56.2 & 0.2 &    $-$34.3 & 0.3 \\
  $-$18.3 &    $-$82.5 & 0.4 &    $-$20.2 & 0.3 \\
  $-$18.4 &    $-$47.0 & 0.7 &     +45.0 & 0.5 \\
  $-$18.3 &    $-$32.9 & 0.5 &     +14.1 & 0.4 \\
  $-$18.7 &    $-$16.5 & 0.3 &     +43.7 & 0.4 \\
  $-$19.1 &    $-$38.5 & 0.9 &      +1.8 & 2.1 \\
  $-$19.7 &     +67.6 & 0.8 &     +12.5 & 0.5 \\
  $-$19.4 &    $-$40.7 & 0.6 &    $-$29.6 & 0.5 \\
  $-$20.2 &     +53.0 & 3.3 &      +6.1 & 1.3 \\
  $-$20.8 &     +45.9 & 1.9 &     $-$9.9 & 0.2 \\
  $-$20.7 &    $-$73.5 & 0.2 &    $-$25.6 & 0.4 \\
  $-$21.3 &    $-$72.7 & 0.3 &     $-$5.9 & 0.3 \\
  $-$20.7 &    $-$52.2 & 0.4 &     +18.1 & 0.4 \\
  $-$21.5 &    $-$37.4 & 1.1 &     +30.1 & 0.2 \\
  $-$21.7 &    $-$32.5 & 0.1 &     +12.8 & 0.1\\
  $-$21.8 &      +1.4 & 1.1 &    $-$21.6 & 0.9 \\
  $-$22.4 &    $-$34.5 & 0.6 &      +0.4 & 0.7 \\
  $-$23.7 &    $-$49.8 & 0.6 &    $-$11.0& 0.4 \\
  $-$24.2 &    $-$50.5 & 0.5 &    $-$27.7 & 0.4 \\
  $-$24.0 &    $-$93.6 & 1.1 &    $-$12.0 & 0.5 \\
  $-$28.2 &     +11.2 & 0.3 &    $-$13.1 & 0.8 \\
  $-$26.1 &     $-$0.3 & 0.6 &      +2.7 & 0.5 \\
  $-$27.8 &     +12.0 & 0.5 &     +54.1 & 0.7 \\
  $-$27.9 &    $-$28.7 & 0.4 &     +37.4 & 0.7 \\
  $-$27.2 &     +14.1 & 0.9 &    $-$32.4 & 0.7 \\
  $-$29.0 &     +27.4 & 0.4 &    $-$11.2 & 0.6 \\
  $-$30.0 &     +23.9 & 1.2 &     +54.9 & 1.0 \\
  $-$33.8\footnotemark[$\ddag$]&     $-$0.1 & 0.9 &     $-$2.0 & 2.7 \\
  $-$32.5 &      +5.4 & 0.1 &     +11.4 & 0.2 \\
  $-$33.3 &    $-$29.4 & 0.4 &     $-$1.8 & 0.3 \\  
\hline
 
\end{tabular*}
\end{center}

\par\noindent
\footnotemark[$*$]Relative R.A. offset.
\par\noindent
\footnotemark[$\dag$]Relative Dec. offset.
\par\noindent
\footnotemark[$\ddag$]Maser feature containing the phase-reference spots.

\end{table}

  
\begin{table}[h]
\begin{center}
\caption{Same as table 2, but at the 2nd epoch (1997 March).}

\begin{tabular*}{\columnwidth}{@{\hspace{\tabcolsep}
\extracolsep{\fill}}cccccc}
\hline\hline 
 {Feature}&
 {$V_{\rm LSR}$ }&
 $\Delta x$\footnotemark[$*$] & Error &
 $\Delta y$\footnotemark[$\dag$] & Error \\
 {}&
 [km s$^{-1}$] &
 [mas] & [mas] &
 [mas] & [mas] \\
 \hline
{\sf a}& $-$13.9 &+37.0   &0.2 &$-$19.3&0.1\\
{\sf b}& $-$16.6 &$-$43.2&0.1 &+56.8   &0.1\\
{\sf c}& $-$17.0 &+11.1   &0.1 &$-$28.8&0.1\\
{\sf d}& $-$21.7 &$-$33.3&0.1 &$-$28.4&0.1\\
$\cdots$    &$-$22.1 &$-$28.7&0.3 &$-$25.2&0.1\\
{\sf e}& $-$23.5 &$-$46.5&0.1 &+3.37   &0.3\\
{\sf f}& $-$29.4 &+17.1   &0.1 &+12.0   &0.1 \\
{\sf g}& $-$30.1 &+16.2   &0.2 &+9.6    &0.2 \\
$\cdots$     &$-$28.9 &$-$3.4 &0.3 &+25.9   &0.2\\
$\cdots$     &$-$29.2 &+24.3   &0.1 &+18.0   &0.1\\
$\cdots$     &$-$29.3 &$-$4.8 &0.2 &+25.4   &0.1\\
{\sf h}& $-$31.2 &+21.2   &0.2 &+9.6    &0.2 \\
{\sf i}& $-$31.3 &+7.5    &0.1 &+3.8    &0.2 \\
$\cdots$     &$-$32.1 &+0.8    &0.3 &+4.5    &0.2\\
{\sf j\footnotemark[$\ddag$]}&$-$33.4&$-$0.1   &0.2&+0.6  &0.2 \\
{\sf k}&$-$34.2 &+3.5    &0.1 &$-$0.4 &0.2\\
\hline
 
\end{tabular*}
\end{center}

\par\noindent
\footnotemark[$*$]Relative R.A. offset.
\par\noindent
\footnotemark[$\dag$]Relative Dec. offset.
\par\noindent
\footnotemark[$\ddag$]See table 2.
\end{table}

  
\begin{table}[h]
\begin{center}
\caption{Same as table 2, but at the 3rd epoch (1997 May).}

\begin{tabular*}{\columnwidth}{@{\hspace{\tabcolsep}
\extracolsep{\fill}}cccccc}
\hline\hline 
 {feature}&
 {$V_{\rm LSR}$ }&
 $\Delta x$\footnotemark[$*$] & Error &
 $\Delta y$\footnotemark[$\dag$] & Error \\
 {}&
 [km s$^{-1}$] &
 [mas] & [mas] &
 [mas] & [mas] \\
 \hline
{\sf a} & $-$14.7& +37.5   &0.2 &$-$21.3&0.2\\
{\sf b} & $-$17.4& $-$45.4&0.3 &+56.2   &0.2\\
$\cdots$   & $-$17.4& $-$41.2&0.2 &+54.0   &0.2\\
$\cdots$   & $-$17.4& +11.2   &0.2 &$-$23.4  &0.4\\
{\sf c} & $-$17.7& +12.1   &0.2 &$-$29.4&0.4\\
{\sf d} & $-$22.5& $-$31.5&0.2 &$-$28.3&0.4\\
$\cdots$   & $-$23.2& $-$17.6&0.2 &+13.8   &0.2\\
$\cdots$   & $-$23.2& $-$21.7&0.2 &+15.4   &0.2\\
{\sf e} & $-$24.4& $-$46.1&0.2 &+1.6    &0.2\\
{\sf f} & $-$30.1& +18.3   &0.2 &+10.7   &0.3\\
{\sf g} & $-$30.9& +16.3   &0.4 &+7.8    &0.2\\
{\sf h} & $-$32.1& +21.3   &0.2 &+7.2    &0.2\\
{\sf i} & $-$32.1& +9.1    &0.2 &+1.5    &0.2\\
$\cdots$   & $-$32.2& +4.5    &0.3 &+4.5    &0.2\\
{\sf j\footnotemark[$\ddag$]} & $-$34.2& +0.0    &0.2 &$-$1.0 &0.3 \\ 
\hline
 
\end{tabular*}
\end{center}

\par\noindent
\footnotemark[$*$]Relative R.A. offset.
\par\noindent
\footnotemark[$\dag$]Relative Dec. offset.
\par\noindent
\footnotemark[$\ddag$]See table 2.
\end{table}


\begin{table}[h]
\begin{center}
\caption{Same as table 2, but at the 4th epoch (1998 April).}
\begin{tabular*}{\columnwidth}{@{\hspace{\tabcolsep}
\extracolsep{\fill}}cccccc}
\hline\hline 
 {Feature}&
 {$V_{\rm LSR}$ }&
 $\Delta x$\footnotemark[$*$] & Error &
 $\Delta y$\footnotemark[$\dag$] & Error \\
 {}&
 [km s$^{-1}$] &
 [mas] & [mas] &
 [mas] & [mas] \\
\hline
 $\cdots$&$-$13.4	&	+17.1	&	0.2	&	$-$34.2	
 &	0.1	\\ 
$\cdots$&$-$13.7	&	+36.9	&	0.4	&	$-$26.5	&	
 0.4	\\ 
$\cdots$&$-$14.0	&	+35.5	&	0.3	&	$-$27.6	&	
 0.1	\\ 
$\cdots$&$-$14.4	&	+34.3	&	0.2	&	$-$27.0	&	
 0.4	\\ 
$\cdots$&$-$13.7	&	+29.6	&	0.1	&	$-$20.1	&	
 0.1	\\ 
{\sf a}&$-$14.1	&	+43.9	&	0.1	&	$-$21.8	&	
 0.1	\\ 
$\cdots$&$-$15.2	&	+4.1	&	0.1	&	+12.6	&	
 0.2\\ 
$\cdots$&$-$16.4	&	+13.3	&	0.1	&	$-$25.4	&	
 0.2	\\ 
{\sf b}&$-$16.7	&	$-$44.9	&	0.1	&	57.3	&	
 0.2	\\ 
$\cdots$&$-$17.4	&	+14.1	&	0.1	&	$-$18.9	&	
 0.1	\\ 
$\cdots$&$-$18.0	&	+5.8	&	0.2	&	+20.1	&	
 0.2	\\ 
$\cdots$&$-$18.0	&	+9.1	&	0.2	&	$-$36.6	&	
 0.1	\\ 
$\cdots$&$-$20.6	&	+3.9	&	0.1	&	$-$34.0	&	
 0.1	\\ 
$\cdots$&$-$21.9	&	$-$32.3	&	0.2	&	$-$25.0	&	
 0.2	\\ 
$\cdots$&$-$28.4	&	+18.6	&	0.1	&	+10.8	&	
 0.2	\\ 
$\cdots$&$-$29.1	&	+4.5	&	0.1	&	+7.1	&	
 0.2	\\ 
$\cdots$&$-$29.1	&	$-$0.9	&	0.2	&	+15.5	&	
 0.1	\\ 
$\cdots$&$-$29.2	&	+23.5	&	0.2	&	+0.1	&	
 0.1	\\ 
$\cdots$&$-$29.3	&	+18.3	&	0.1	&	+8.9	&
 0.1	\\
$\cdots$&$-$29.5	&	$-$6.4	&	0.1	&	+24.0	&	
 0.1	\\ 
$\cdots$&$-$30.4	&	+17.9	&	0.3	&	+7.4	&	
 0.1	\\ 
$\cdots$&$-$30.5	&	+4.7	&	0.2	&	+6.9	&	
 0.2	\\ 
$\cdots$&$-$31.3	&	+7.2	&	0.2	&	$-$3.5	&	
 0.2	\\ 
$\cdots$&$-$32.0	&	$-$71.3	&	0.1	&	$-$50.0	&	
 0.1	\\ 
$\cdots$&$-$32.8	&	$-$43.8	&	0.2	&	$-$3.9	&	
 0.1	\\ 
$\cdots$&$-$32.4	&	+12.1	&	0.1	&	+0.2	&	
 0.3	\\ 
{\sf j\footnotemark[$\ddag$]}&$-$33.5	&	+0.2	&	0.2	
 &	+0.4	&	0.1	\\ 
{\sf k}&$-$34.4	&	+3.7	&	0.1	&	$-$0.5&	
 0.2	\\ 
$\cdots$&$-$35.3	&	+2.5	&	0.1	&	$-$5.2	&	
 0.2	\\ 
\hline
\end{tabular*}
\end{center}

\par\noindent
\footnotemark[$*$]Relative R.A. offset.
\par\noindent
\footnotemark[$\dag$]Relative Dec. offset.
\par\noindent
\footnotemark[$\ddag$]See table 2.
 
\end{table}


\begin{table}[h] 
\begin{center}
\caption{Parameters of the proper motions of the maser features in IRC+60169.}
\begin{tabular}{ccccc}
\hline\hline
 Feature&
 $\mu_{\rm x}$\footnotemark[$*$]&
 Error &
 $\mu_{\rm y}$\footnotemark[$\dag$]&
 Error\\
 {}&
 [mas yr$^{-1}$]&
 [mas yr$^{-1}$]&
 [mas yr$^{-1}$]&
 [mas yr$^{-1}$]\\
 \hline
 {\sf a}&	+0.3	&	0.5	&	$-$3.0	&	0.3\\
 {\sf b}&	$-$8.2	&	0.2	&	+0.0	&	0.4\\
 {\sf c}&	+3.5	&	0.4	&	+5.9	&	0.6\\
 {\sf d}&	+8.8	&	0.3	&	+9.3	&	0.6\\
 {\sf e}&	+0.1	&	0.3	&	$-$1.2	&	0.5\\
 {\sf f}&	+5.1	&	0.3	&	+1.2	&	0.4\\
 {\sf g}&	$-$1.7	&	0.6	&	$-$1.5	&	0.4\\
 {\sf h}&	$-$1.7	&	0.4	&	$-$4.9	&	0.4\\
 {\sf i}&	+7.1	&	0.3	&	$-$5.1	&	0.3\\
 {\sf j}&	$-$6.8	&	0.3	&	$-$0.7	&	0.3\\
 {\sf k}&	$-$6.5	&	0.2	&	$-$0.6	&	0.4\\
\hline
\end{tabular}
\end{center}
 
\par\noindent
\footnotemark[$*$]Motions in the R.A. direction.
\par\noindent
\footnotemark[$\dag$]Motions in  the Dec. direction.
\end{table}


\begin{figure}
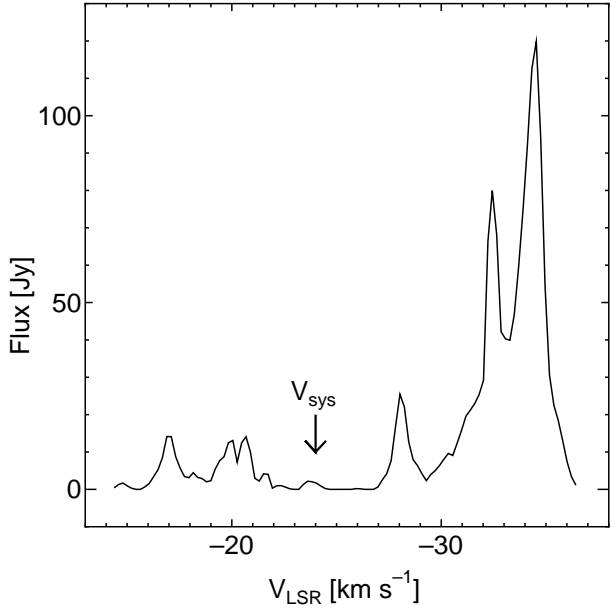

   \begin{center}
      \FigureFile(80mm,80mm){figure1.epsi}
   \end{center}
   \caption{Cross-correlated spectrum of the water maser emission of
 IRC+60169, obtained by the Kashima--Nobeyama baseline at the 1st
 epoch. $V_{\rm sys}$ is the stellar systemic velocity of IRC+60169. }  
\end{figure}


\begin{figure*}[h]
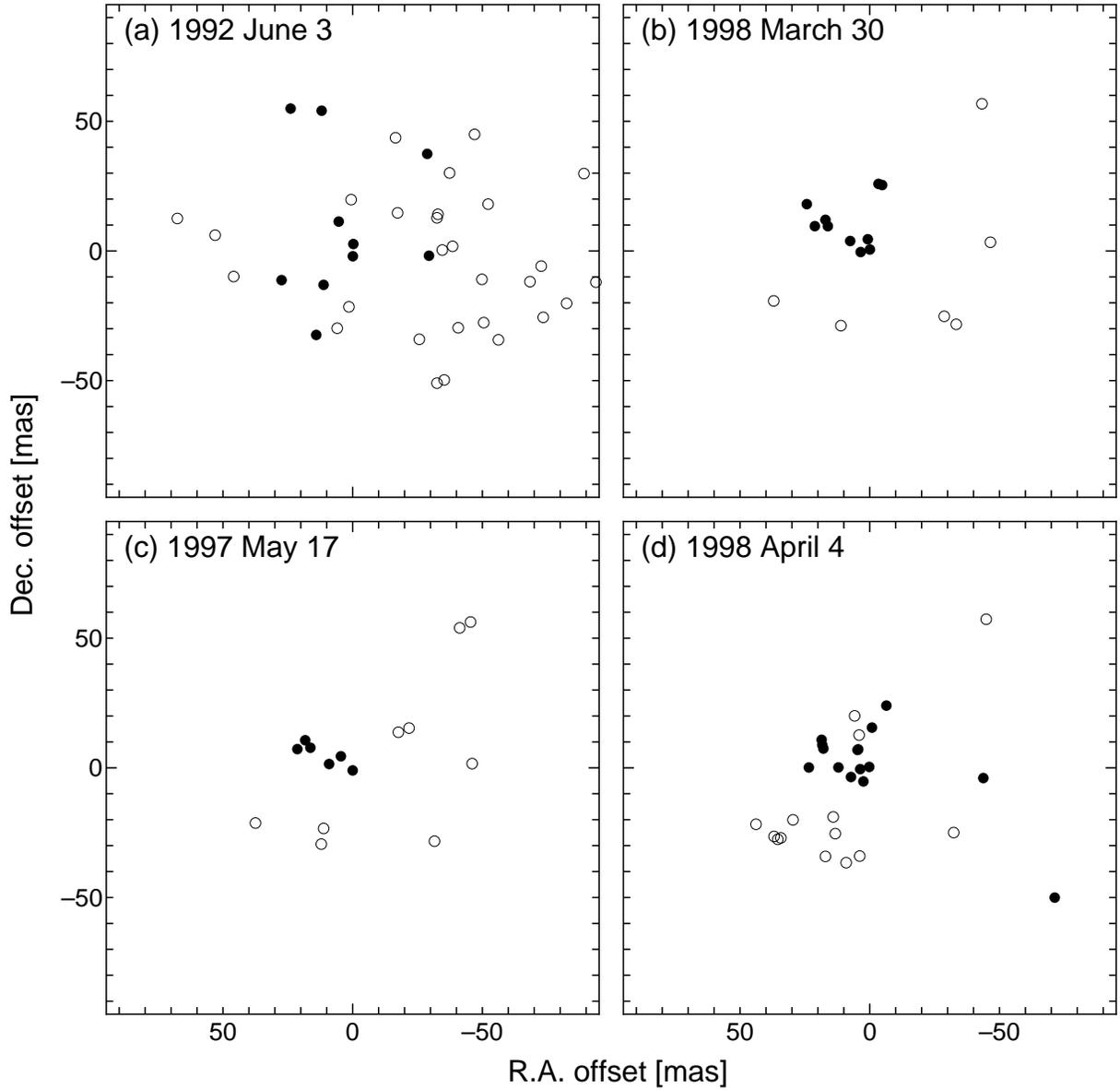

   \begin{center}
      \FigureFile(160mm,160mm){figure2.epsi}
   \end{center}
   \caption{Distributions of the water maser features of IRC+60169
 on (a) 1992 June  3, (b)  1997 March 30, (c)
 1997 May 7, and (d) 1998  April 4. The 
 open and filled circles indicate the maser features of the B and R components,
 respectively. } 
\end{figure*}


\begin{figure}
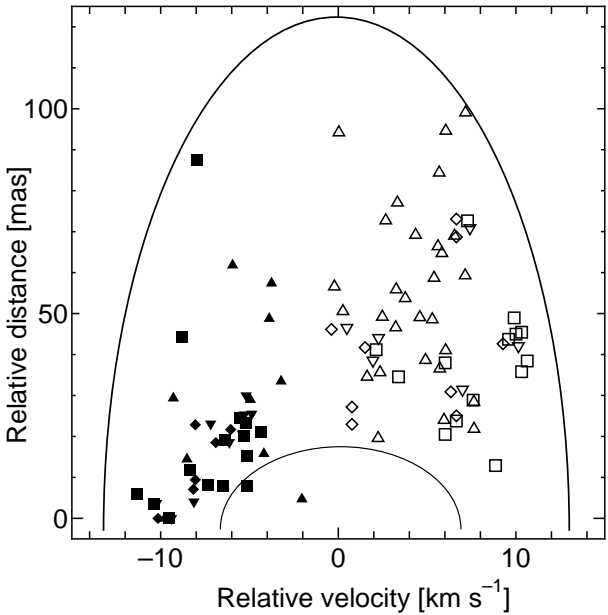

   \begin{center}
      \FigureFile(80mm,80mm){figure3.epsi}
   \end{center}
   \caption{Radial velocities of the maser features relative to the stellar
 velocity ($V_{\rm LSR} = -24$ km s$^{-1}$) plotted against the distances from 
 the feature {\sf j}. The filled and open marks indicate the B and R
 components, respectively. The triangles, inverse triangles,
 diamonds, and 
 squares show the maser features at the 1st, 2nd, 3rd, and 4th
 epochs, respectively. } 
\end{figure}


\begin{figure}
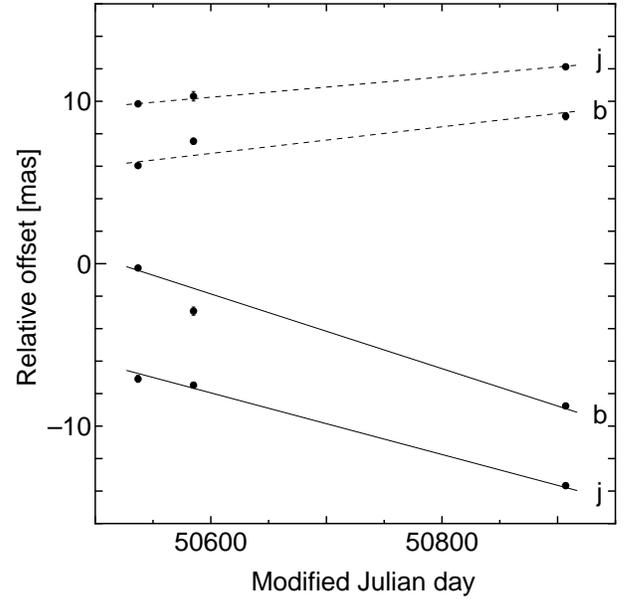

   \begin{center}
      \FigureFile(80mm,80mm){figure4.epsi}
   \end{center}
   \caption{Proper motions of the maser features {\sf b} and {\sf
 j}.  The solid
 line indicates a lest-squares fitted line assuming a constant velocity
 in the R.A. direction. The dotted line indicates
 that in the Dec. direction. } 
\end{figure}


\begin{figure}
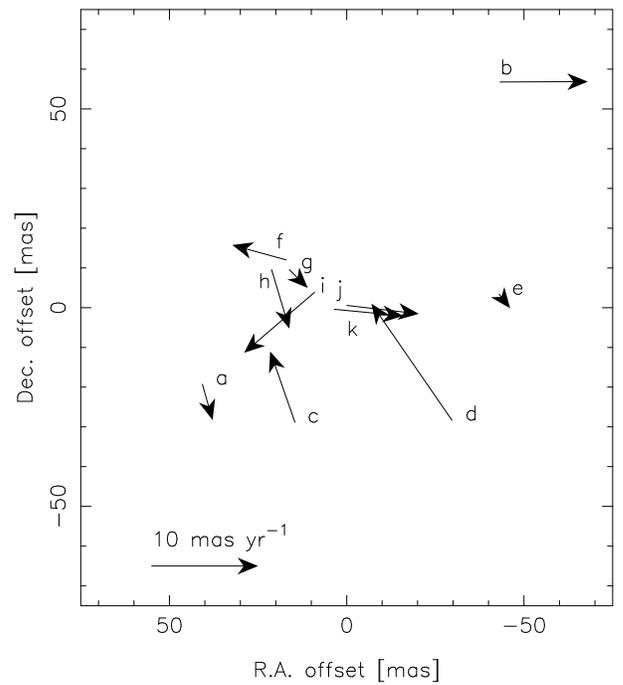

   \begin{center}
      \FigureFile(80mm,80mm){figure5.epsi}
   \end{center}
   \caption{Proper motions of the water maser features. We 
calculated the average of  all of the proper-motion vectors and 
subtracted it from each proper-motion vector. } 
\end{figure}


\begin{figure}
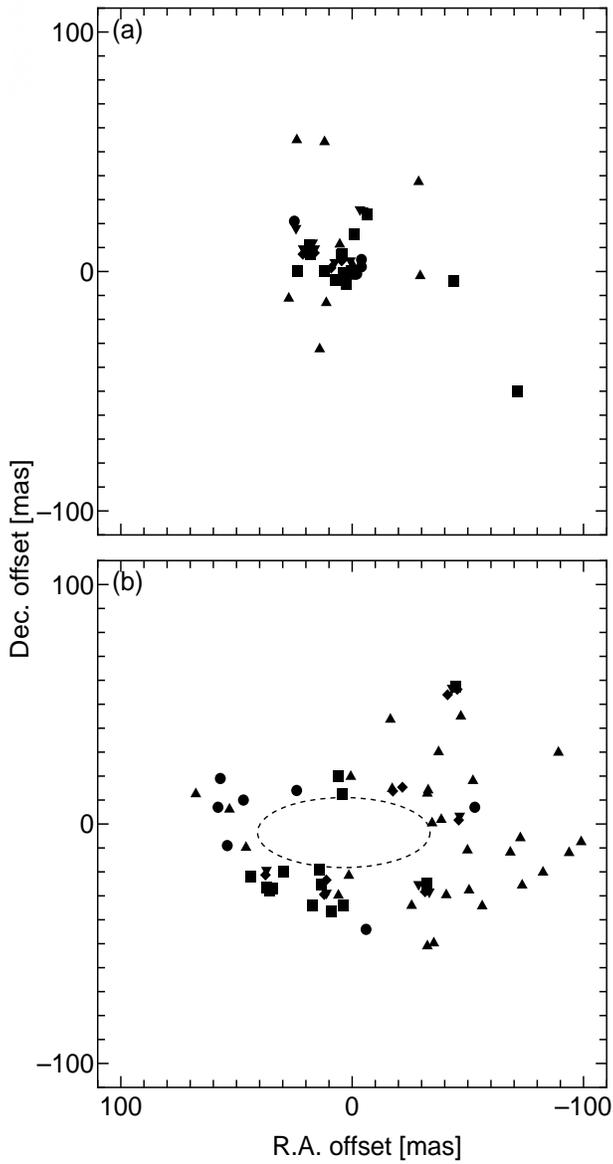

   \begin{center}
      \FigureFile(80mm,80mm){figure6.epsi}
   \end{center}
   \caption{Superposition of the maser feature distributions between
 1990 and 
 1998 of (a) the B components and (b) the  R components. The filled
 circles, triangles, inverse triangles, lozenges, and 
 squares show the data of Colomer et  al. (2000), the 1st, 2nd,
 3rd, and 4th epochs of the present work, respectively. The dotted
 ellipse indicates the putative blocking region.} 
\end{figure}


\begin{figure}
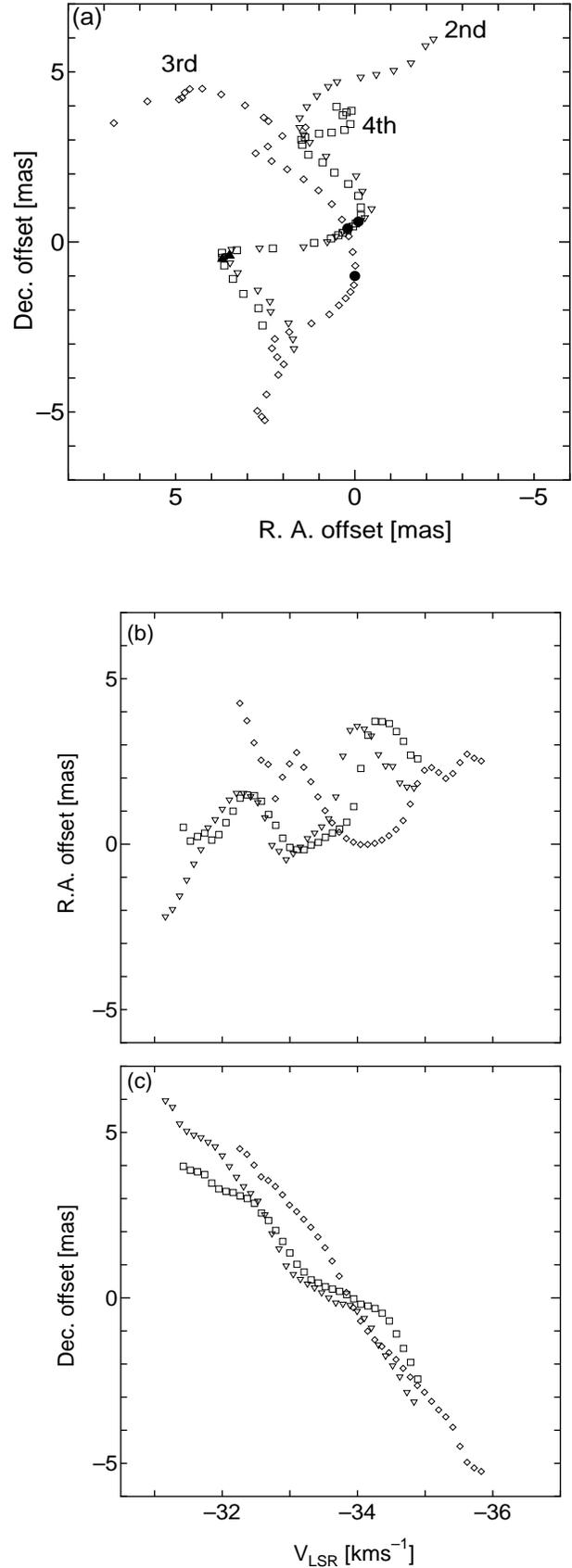

   \begin{center}
      \FigureFile(80mm,80mm){figure7.epsi}
   \end{center}
   \caption{Alignment change of the water maser spots in the
 features {\sf j} 
 and {\sf k}, (a) R.A. vs. Dec., (b) $V_{\rm LSR}$ vs. R.A., and (c)
 $V_{\rm LSR}$ vs. Dec. The
 open inverse triangles, open diamonds, and open squares indicate the
 positions of the spots at the 2nd, 3rd, and 4th epochs,
 respectively. The filled circle and filled triangle show features {\sf j} 
 and {\sf k}, respectively. }  
\end{figure}


\end{document}